\documentstyle[11pt,paspconf,epsf]{article}

\begin{document}

\title{Synthetic stellar mass-to-light ratios for stellar
populations}

\author{Claudia Maraston}
\affil{Dipartimento di Astronomia, Universit\`a di Bologna, Italia}

\begin{abstract}
Evolutionary synthesis models for stellar populations of various ages and 
chemical compositions are constructed with the approach described in Maraston 
(1998), in which the Fuel Consumption Theorem is used to evaluate the 
energetics of Post Main Sequence stars. We present here the synthetic 
``{\it stellar}" mass-to-light ratios ($M^{*}/L$) in the $U,B,V,R,I,J,H,K$ 
photometric bands, as functions of age and chemical composition, for single 
burst populations. Taking into account the contribution by stellar dead 
remnants, the computed $M^{*}/L$ ratios can be directly compared to those
measured in early-type galaxies. The dependence of $M^{*}/L$ ratios on the IMF
slope is also explored. The most interesting result is that the 
$M^{*}/L_{B}$ ratio of a 15 {\rm Gyr} stellar population is found to 
increase by nearly a factor of three, when the chemical composition rises from
 ${\rm[Fe/H]} \sim -0.5$ to ${\rm[Fe/H]} \sim +0.3$.\ This impacts on the 
 interpretation of the tilt of the {\it Fundamental Plane} of cluster ellipticals in the $B$ 
band.
\end{abstract}


\section{Introduction}

Elliptical galaxies (Es) are not randomly distributed in the three-dimensional 
space defined by the observed parameters: central velocity dispersion 
$\sigma_0$, effective radius $r_{\rm e}$ and the surface brightness $I_{\rm e}$
 within $r_{\rm e}$.\ They occupy with very small dispersion a planar surface 
 (Dressler {\it et al.}\ 1987; Djorgovski and Davis 1987), called the {\it Fundamental
  Plane} (FP). 
Thus the FP defines a correlation between the luminosity, size and dynamical 
mass of a galaxy. With the aim of a more transparent view on the physical 
properties of Es, Bender, Burstein \& Faber (1992; hereafter BBF) analysed the 
Virgo and Coma cluster elliptical FP in the $B$-band, by introducing a
convenient orthogonal coordinate system (known as the {\it ${\kappa}$-space}), 
in which the new variables are linear combinations of 
$\log{{\sigma_0}^2}$, $\log{R_{\rm e}}$ and $\log{I_{\rm e}}$.
In this frame, the $\kappa_1$-$\kappa_3$ plane is an almost edge-on view of the
 FP (see BBF, Fig.~1). The main properties of the BBF Es FP are 
1) the so-called {\it tilt}, i.e.\ the systematic increase of $\kappa_3$ along
the FP ($\kappa_3=0.15\kappa_1+const$) and 2) its tightness, i.e.\ the very 
small dispersion of $\kappa_3$, ${\sigma({\kappa_3})}\simeq\pm0.05$ at every 
location on the FP. Using the virial theorem, the ${\kappa}$-coordinates can be 
related to the total galaxy mass, $M=c_{2}r_{\rm e}{\sigma_0}^{2}$, by 
$\kappa_1\propto \log {M/c_{2}}$ and $\kappa_3\propto \log M/{L_{B}c_{2}}$.
 If the virial coefficient $c_{2}$ is constant for all the galaxies (or, Es 
 form an {\it homologous} family), the observed FP tilt implies that $M/L_{B}$ 
 increases by a factor $\sim$ 3 with galactic mass. 
The physical origin of the FP tilt is still at the debate (see also Pahre {\it 
et al.},\ this volume), but it can be sought in two {\it orthogonal} directions:
 either it is due to a {\it stellar population} effect, in which case $M^{*}/L$
 scales with the luminosity (mass) of the galaxies and $c_{2}$ is constant. Or 
 {\it structural/dynamical} homology is broken, in which case $M^{*}/L$ is 
 constant and $c_{2}$ varies along the galaxy sequence.  
 The aim of this contribution is to explore the former option, the 
 structural/dynamical one being extensively discussed in, e.g., Ciotti, 
 Lanzoni \& Renzini (1996). 
 
From an observational point of view, the slope ${\alpha}$ of the FP scaling 
relation, $r_{\rm eff} \propto {\sigma_0^{\alpha}}{I_{\rm e}^{\beta}}$, is 
found to increase systematically with wavelenght, from the $U$-band to the 
$K$-band (see the contribution by Pahre {\it et al.}\ in this volume and 
references therein). 
This points towards systematic variations of the stellar content 
along the Es sequence, which also gets support from systematic trends in 
colours and line strenghts with galaxy luminosity (hence with ${\kappa_1}$). 
   
 The mass-to-light ratios of stellar populations change with age, chemical 
 composition, and Initial Mass Function (IMF). The IMF effect has been explored 
 by Renzini \& Ciotti (1993). 
 Their main conclusion is that a strong {\it fine tuning} is required to 
 produce the observed tilt of the FP, preserving the constant thickness: the 
 IMF should be virtually universal for a given galaxy mass, and yet exhibit a 
 large trend with galaxy mass.
A straightforward interpretation of the ${\rm Mg_{2}}-\sigma$ relation (see 
Bender, Burstein \& Faber 1993) and the colour-magnitude relation (Bower, Lucey
 \& Ellis 1992; see also Kodama, this volume) of Es is in terms of 
 increasing metallicity along the galaxy sequence. 
This effect on $M^{*}/L$ is explored for the $V$-band in Renzini (1995).  
 In this work, the effect of metallicity and age on the $M^{*}/L$ ratio is 
 analysed for several photometric bands.
 This is part of a more extended work which aimes at determining {\it what 
 fraction of the observed FP tilt is produced by stellar population effects},
 through the comparison of the FP at different wavelenghts (Maraston \& Renzini
 1998, {\it in preparation}). 
   
\section{Models for stellar mass-to-light ratios}

Models for Simple Stellar Populations (SSP) are computed with the
evolutionary synthesis approach described in Maraston (1998; hereafter M98),
in which the Fuel Consumption theorem (Renzini \& Buzzoni 1986) is used to 
evaluate the energetic of Post Main Sequence (PMS) stars. 

The input stellar tracks for the SSP models presented here are taken from 
Cassisi ({\it private communication}; see also Bono {\it et al.}\ 1997). They are 
computed by means of the FRANEC evolutionary code (see Chieffi \& Straniero 
1989). The range in metallicity is ${\rm [Fe/H]}=[-1.3\div0.3]$, with 
helium-enrichment ${\rm\Delta Y/\Delta Z}$=2.5; age varies from 3 {\rm Gyr} to
 15 {\rm Gyr}. We refer to Maraston (1998, {\it in preparation}) for more 
 details.

The IMF is assumed to follow the usual power-law, $\Psi(m)=A{m}^{-(1+x)}$. 
Besides the Salpeter slope ($x$=1.35), we explore the multislope case with a 
break at low masses, as recently suggested by Gould, Flynn \& Bahcall (1997; 
hereafter GBF97) on the basis of HST observations of disk $M$ dwarfs. Depending
 on the mass range, the adopted slopes are: $x$=0 for $ m \leq 0.6 M_{\odot}$, 
 $x$=1.21 for $ 0.6 < m/M_{\odot}\leq 2 $ and $x$=1.35 otherwise. The upper and
  lower mass cutoffs are fixed at 100 $M_{\odot}$ and 0.1 $M_{\odot}$ 
  respectively in both cases.
The synthetic $M^{*}/L_{\lambda}$ ratios contain the contribution by 
stellar remnants. As a stellar population ages, stars progressively die, 
leaving remnants, that only contribute to the mass of the population. The mass 
lost in the form of ejected gas is supposed to be blown out of the stellar 
population.
The total mass of an SSP of age $t$, $M^*(t)$, is thus obtained by convolving 
the stellar mass $m^*(t)$ with the IMF, with $m^*(t)=m_{\rm in}$ for 
$m_{\rm in} \leq m_{\rm TO}(t)$ and $m^*(t)=m_{\rm R}$ for 
$m_{\rm in} > m_{\rm TO} (t)$. 
The quantities $m_{\rm in}$, $m_{\rm TO}(t)$ and $m_{\rm R}$ denote the initial
Main Sequence mass, the turnoff mass and the remnant mass, respectively.
For the remnant mass we adopt the same recipe as in M98 and Renzini \& Ciotti
(1993).
We do not take the dependence of $m_{\rm R}$ on metallicity into account, since
 it implies difference in $m_{\rm R}$ of the order of $\sim 10^{-2} M_{\odot}$.

At fixed age, $M^*(t)$ is fairly insensitive to the metal content because 
$ m_{\rm TO} $(t) does not change appreciably with the chemical composition: 
for a 15 Gyr SSP, the variation of $ m_{\rm TO} $(t) is at most 0.15 
$M_{\odot}$.\ The $M^*(t)$ computed with a Salpeter IMF is $\sim$ 2.5 times 
greater than the one for the GBF97 IMF, since the latter contains less low-mass
 stars. Due to the slow evolution of $ m_{\rm TO}(t)$ at old ages ($t \ga$ 1 
 {\rm Gyr}), the influence of age on $M^*(t)$ is very mild (see M98 for more details).
 In the age range $3\div15$ {\rm Gyr}, $M^*(t)$ varies only by a factor 
 of $\sim 1.11$, independent of the chemical composition. 

\subsection{The metallicity effect}

\begin{figure}[ht]
 \plotone{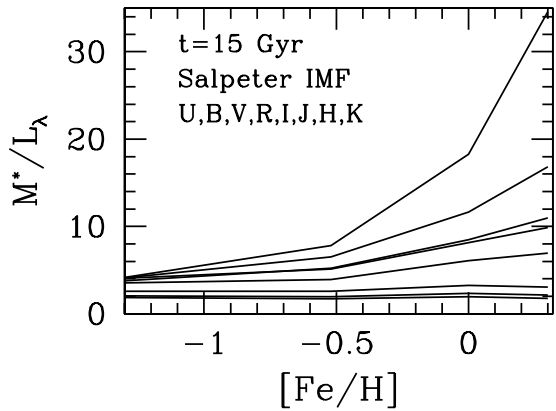}
 \caption{Synthetic $M^{*}/L_{\lambda}$ in $U$,$B$,$V$,$R$,$I$,$J$,$H$,$K$ 
 photometric bands (from top to bottom), as functions of chemical 
 composition.}\label{mlgen}. 
\end{figure}
Fig.~\ref{mlgen} shows the effect of metallicity on mass-to-light ratios. 
The $M^{*}/L_{\lambda}$ ratio for various photometric bands (from $U$ to $K$, 
from top to bottom) as a function of metallicity (for 15 {\rm Gyr}, Salpeter 
IMF SSPs) is shown. In general, the $M^{*}/L_{\lambda}$ ratio increases with
 increasing metallicity. This pattern is more prominent in the $U$-band and
 decreases systematically with increasing wavelenght, leading to a basically
 flat function in the $J,H,K$ bands. 
\begin{figure}[ht]
 \plottwo{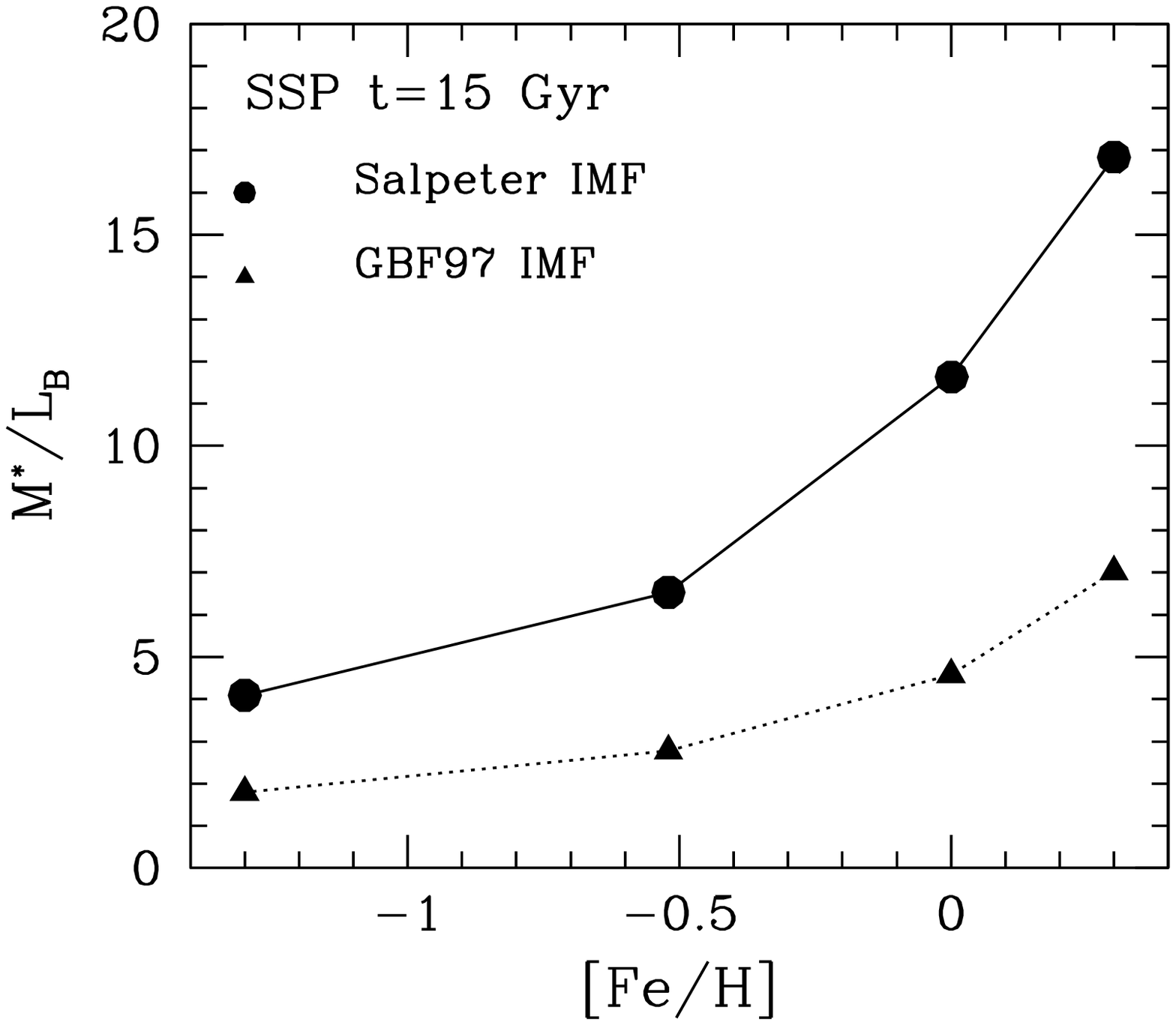}{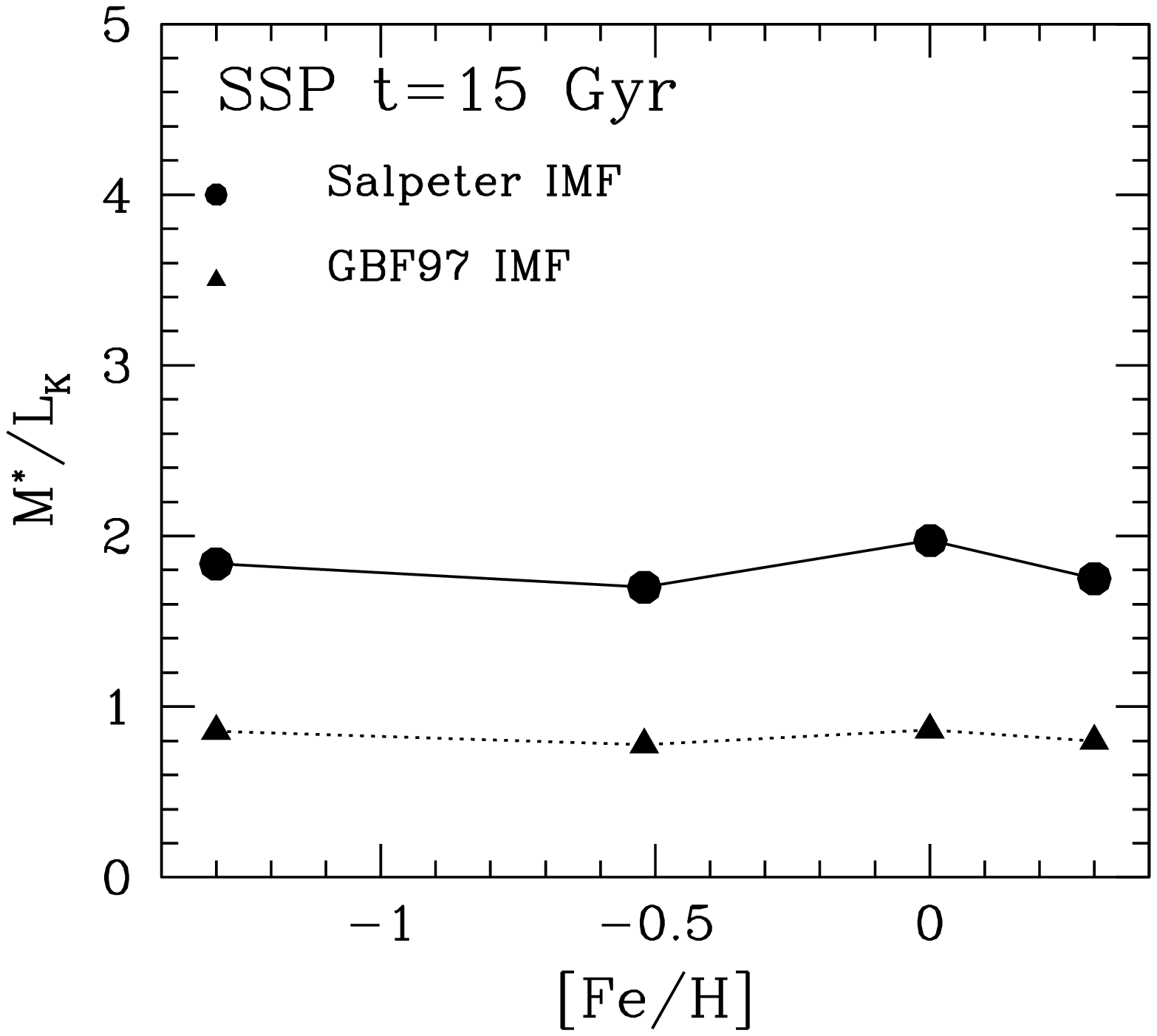}
 \caption{Synthetic $M^{*}/L_{B}$ ({\it left-hand panel}) and $M^{*}/L_{K}$ 
 ({\it right-hand panel}) for 15 {\rm Gyr} SSPs, as 
 functions of chemical composition. Solid line is for a Salpeter IMF, dotted 
 line for the GBF97 IMF (see the text).}\label{mlconz}. 
\end{figure}
Fig.~\ref{mlconz} shows the systematic trend induced by metallicity on 
$M^{*}/L_{B}$ and $M^{*}/L_{K}$ (left-hand and right-hand panel, 
respectively), for the two different IMFs adopted. $M^{*}/L_{B}$ is found 
to increase by a factor $\sim3$ when ${\rm [Fe/H]}$ increases from 
$\sim 1/3 $ solar to two times solar. This range in metal content is roughly 
consistent with the one spanned by BBF Es (see Renzini 1995 and references 
therein), and the BBF FP in the $B$-band is tilted in a way that implies a 
$ M/L_B$ variation of $\sim$ 3 {\it if the assumption of homology holds} 
(see Sec. 1.).
A {\it pure metallicity effect} thus suffices to explain the FP tilt observed 
in the $B$-band. If the low bright Es of BBF sample have ${\rm [Fe/H]}$ 
$\sim-0.3$ (half-solar) then the metallicity effect explains $\sim68 \%$ of 
the observed tilt. These results are for $\Delta Y/\Delta Z$=2.5; it remains to
 be investigated if this is appropriate for the BBF Es (see Renzini 1995).  
To assess what fraction of the FP tilt is due to metallicity effects, the trend
 of the tilt with wavelenght has to be examined.
In fact, in the whole metallicity range here considered, the $M^{*}/L$ ratios 
in the IR are nearly constant (see Fig.~\ref{mlconz}, right-hand panel for the 
$K$ band).
This scope will be pursued in a future paper (Maraston \& Renzini, {\it in 
 preparation}).  

Fig.~\ref{mlconz} shows that the two IMFs considered here manifest themselves 
only in the absolute values of $M^{*}/L_{\lambda}$, the systematic trends
discusses above being not affected. This follows from $m_{TO}$ being 
fairly independent on the chemical composition (see Maraston 1998, {\it in 
preparation} for a wider discussion). 

\subsection{The age effect}

\begin{figure}[ht]
 \plottwo{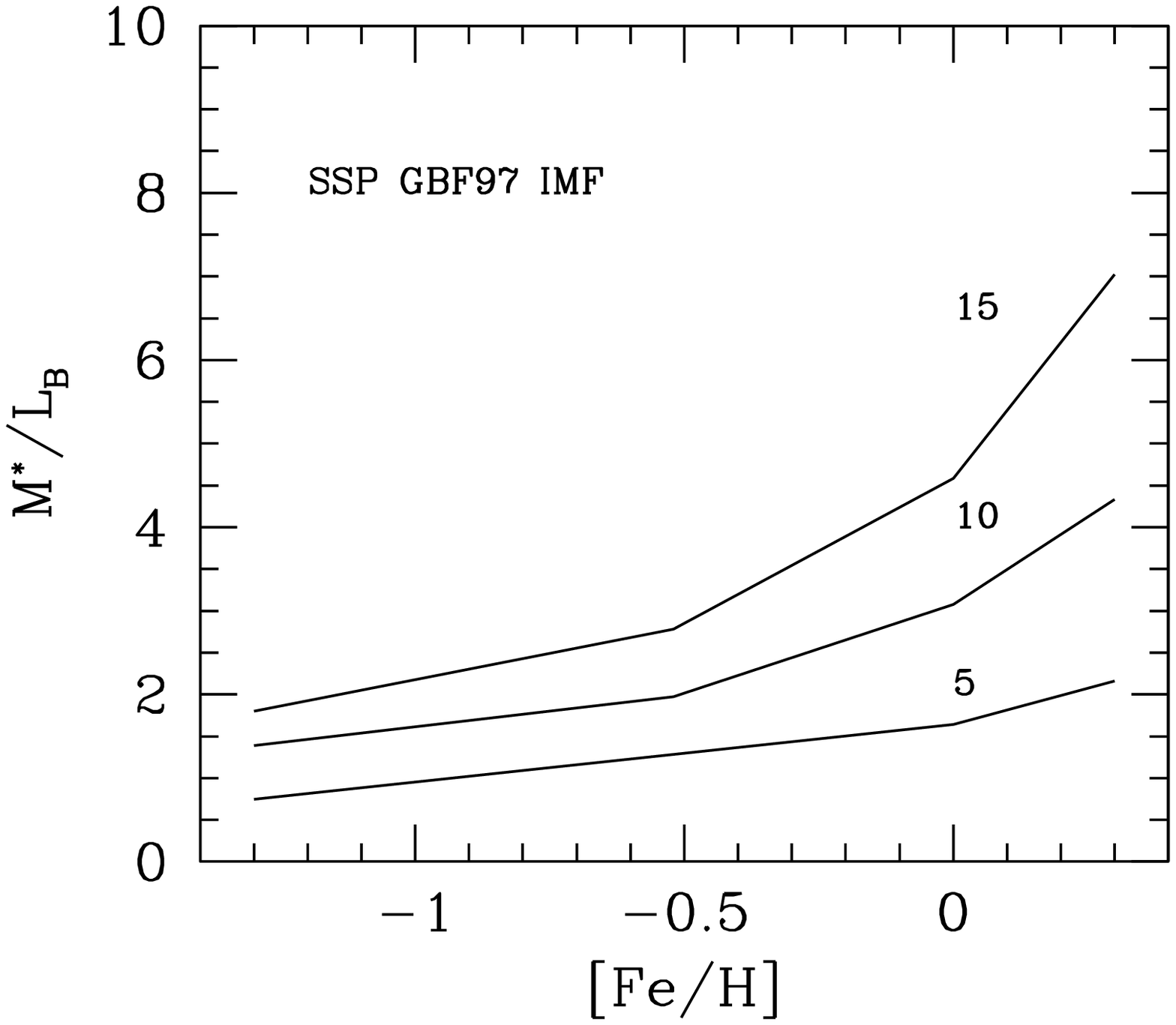}{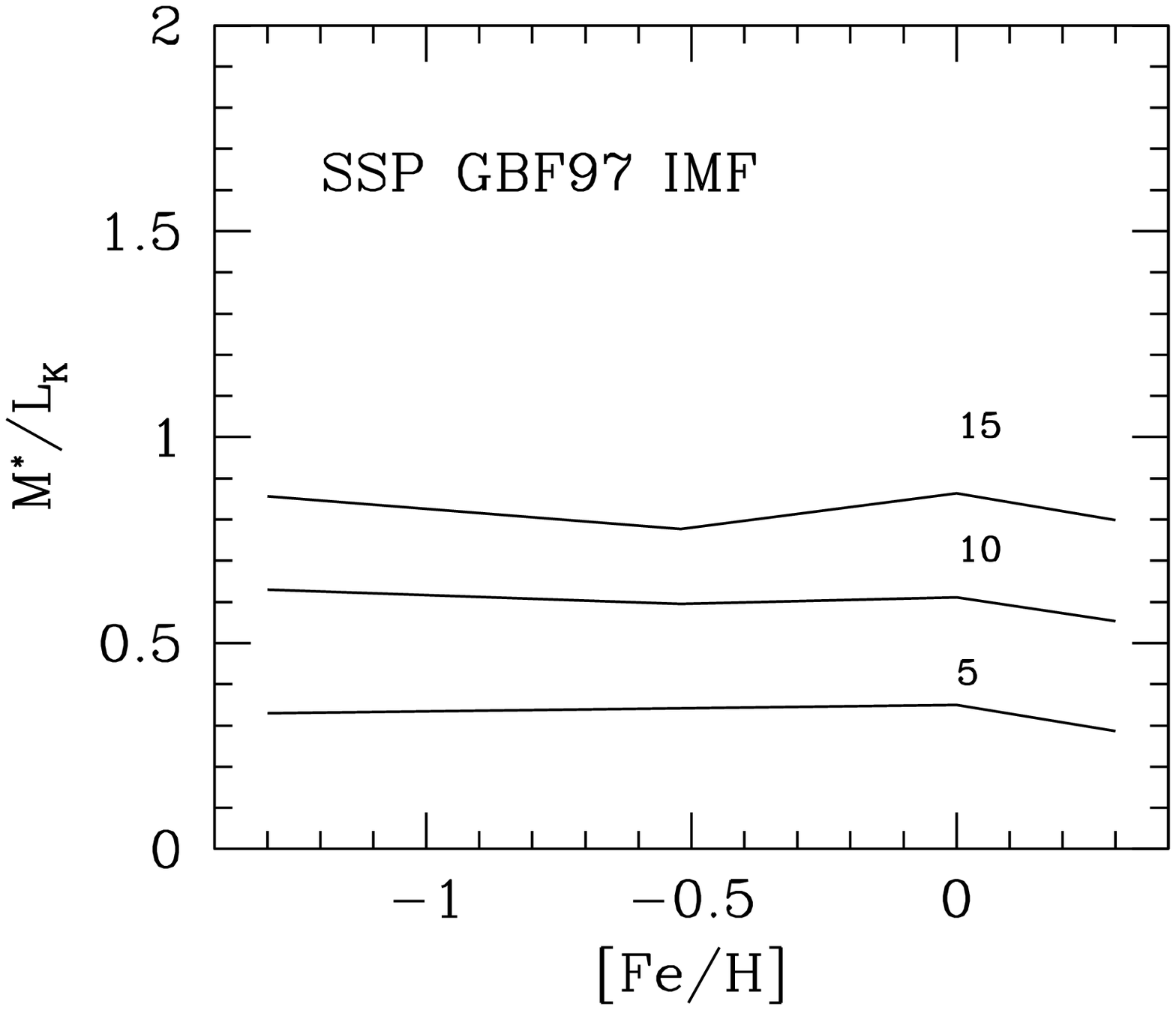}
 \caption{$M^{*}/L_{B}$ ({\it left-hand panel}) and $M^{*}/L_{K}$ 
 ({\it right-hand panel}) as functions of metal content, for 3, 10
 and 15 {\rm Gyr} SSPs. Models are for GBF97 IMF (see the text).}
 \label{mlconage}
\end{figure}
Besides metallicity, age induces a systematic trend in the mass-to-light ratio. 
Here we briefly comment on this point. Fig.~\ref{mlconage} shows the age effect
 on $M^{*}/L$ in the $B$ and $K$ bands (left-hand and right-hand panel, 
 respectively) for the GBF97 IMF. For the fixed solar metallicity 
 ${\rm [Fe/H]}$=0, a variation by a factor of $\sim $ 3 in $M^{*}/L_{B}$ is 
 achieved, if age increases from $\sim$ 5 {\rm Gyr} to $\sim$ 15 {\rm Gyr} 
 along the galaxy sequence. 
In this same age range, $M^{*}/L_{K}$ varies by a factor of $\sim 2.6$. 
To accept the age solution, a highly syncronisation in the formation process of
 the Es of given luminosity (mass) is required, in order to preserve the small
  FP scatter (see Sec. 1.).  

\section{Conclusions}

We have explored the systematic trends induced by metallicity and age on 
synthetic mass-to-light ratios, in which the contribution by stellar remnants 
is taken into account. The main result is that a pure metallicity effect is 
capable to explain the variation of $M^{*}/L_{B}$ observed in the BBF sample of
 cluster ellipticals, if ${\rm [Fe/H]}$ increases from $\sim -0.5$ to
 $\sim+0.3$ along this galaxy sequence. $M^{*}/L$ ratios in the IR bands are 
instead predicted to display a flat trend with the chemical composition (cf. 
 Fig.~\ref{mlconz}). 

The slope of the FP is found to increase with wavelenght (Pahre {\it et al.},\ 
this volume): the exponent $\alpha$ of the FP standard form 
$r_{\rm eff} \propto {\sigma_0^{\alpha}}{I_{\rm e}^{\beta}}$ is $\sim 1.4$ for 
the BBF FP in the $B$-band and $\sim 1.53$ for the Pahre {\it et al.}\ FP in 
the IR. The 
amplitude of the increase is thus modest, and a puzzling situation appears, in 
which other effects, like non-homology, have to be assumed to work 
differentially with wavelenght, which obviously should not be the case. 
Moreover, a part of the tilt is certainly due to metallicity effects, as 
chemical composition varies along the Es sequence. We find indeed that these 
effects explain $\sim 70-100\%$ of the FP tilt in the $B$-band, depending on 
the real metallicity range spanned by the BBF sample. It is interesting to
mention, however, that the tilt of the IR FP in the {\it ${\kappa}$-space}
derived by Pahre {\it et al.}\ (1998) does not deviate from the BBF one in the
$B$-band (see Sec. 1.).
     
If the IR FP is tilted in the way shown by Pahre {\it et al.},\ then it is 
entirely due to effects other than metallicity. But these effects have to show 
sign also in the $B$-band.   

It would be highly useful measuring the FP correlations in different bands  
{\it for the same sample of galaxies} and using consistent fitting procedures.

\acknowledgments

The author is indebted with Santi Cassisi, for kindly providing the stellar 
tracks in advance of publication, and with Alvio Renzini, Laura Greggio, Ralf 
Bender and Michael Andrew Pahre for many interesting discussions. A special 
thank to Daniel Thomas.

\begin{question}{ROCCA-VOLMERANGE}
The giant branch of globular clusters is known to be very sensitive to
metallicity, how can you explain the constancy of $M^{*}/L_{\rm K}$ when 
$L_{\rm K}$ is dominated by giant branches? 
\end{question}
\begin{answer}{MARASTON}
This effect is the result of the conspiracy between the amount of total 
energy at an SSP disposal and the effective temperatures at which this 
energy is emitted. For $\Delta Y/\Delta Z \sim 2.5$, passing from 
${\rm [Fe/H]}\sim -0.5$ to ${\rm [Fe/H]}\sim +0.3$, the specific fuel burned in 
 the Red Giant Branch by an SSP decreases by a factor $\sim$ 1.32. The reason 
 of this is simply that the percentage in mass of hydrogen, that is always the 
 major source of energy, decreases from $\sim$ 0.75 to $\sim$ 0.62, for 
 $\Delta Y/\Delta Z \sim 2.5$ and the values of ${\rm [Fe/H]}$ here considered. 
 But the fuel burned is released at cooler effective temperatures in a metal 
rich population. Thus the infrared emission is not too much metal dependent.
\end{answer}


\begin{references}
\reference Bender, R., Burstein, D., Faber, S. M. 1992, \apj, 399, 462
\reference Bender, R., Burstein, D., Faber, S. M. 1993, \apj, 411, 153
\reference Bono, G., Caputo, F., Cassisi, S., Castellani, V. \& Marconi, M.
1997, \apj, 489, 822
\reference Bower, R.G., Lucey, J.R. \& Ellis, R. S. 1992, \mnras, 254, 613 
\reference Chieffi, A. \& Straniero, O. 1989, \apjs, 71, 47
\reference Ciotti, L., Lanzoni, B. \& Renzini, A. 1996, \mnras, 282,1
\reference Djorgovski, S., \& Davis, M.  1987, \apj, 313, 59
\reference Gould, A., Bahcall, N. J. \& Flynn, C. 1997, \apj, 482, 913
\reference Pahre, M. A., Djorgovski, S. \& de Carvalho, R.R. 1998, \aj, in
press (astro-ph/9806315)
\reference Maraston, C. 1998, \mnras, in press, (astro-ph/9807338)
\reference Renzini, A. in Stellar Populations, P.C. van der Kruit and G. 
Gilmore (eds.), 1995 IAU, 325
\reference Renzini, A. \& Buzzoni, A. in Spectral Evolution of Galaxies, C.
Chiosi and A. Renzini (eds.), Dordecht, Reidel, 1986, 195
\reference Renzini, A., \& Ciotti 1993, \apjl, 416, L49
\end{references}
\end{document}